\input harvmac
\def \Sc {Schwarzschild}
\def \up {\uparrow}

\def \Q{{\cal Q}}

\def \b {\beta}
\def \chi {\chi}

\def \m {\mu}
\def \n {\nu}

\def \inv {^{-1}}
\def \ov {\over }

\def \lr { \lref}
\def\np {{  Nucl. Phys. }}
\def \pl {{  Phys. Lett. }}
\def \mpl {{ Mod. Phys. Lett. }}
\def \prl {{  Phys. Rev. Lett. }}
\def \pr  {{ Phys. Rev. }}

\def \cqg {{ Class. Quant. Grav. }}

\baselineskip8pt
\Title{
\vbox
{\baselineskip 6pt{\hbox{CERN-TH/96-324}}{\hbox
{Imperial/TP/96-97/07}}{\hbox{hep-th/9611111}} {\hbox{
  }}} }
{\vbox{\centerline {On the structure of composite 
 }
\vskip4pt
 \centerline {black p-brane
configurations and related  black holes}
}}
\vskip -20 true pt
\medskip
\medskip
\centerline{   A.A. Tseytlin\footnote{$^{\star}$}{\baselineskip8pt
e-mail address: tseytlin@ic.ac.uk}\footnote{$^{\dagger}$}{\baselineskip8pt
On leave  from Lebedev  Physics
Institute, Moscow.} }

\smallskip\smallskip
\centerline {\it   Theory Division, CERN, CH-1211  Geneve 23,
Switzerland}
\smallskip\smallskip
\centerline {\it  and  }
\smallskip\smallskip
\centerline {\it Blackett Laboratory, 
Imperial College,  London,  SW7 2BZ, U.K. }

\bigskip\bigskip
\centerline {\bf Abstract}
\medskip
\baselineskip10pt
\noindent
We comment on the structure of intersecting black p-brane 
solutions in string theory explaining how known solutions can
be obtained from Schwarzschild solution simply by sequences
of boosts and dualities. This implies, in particular, that  
dimensional reduction in all internal world-volume directions 
including time leads to a metric (related by analytic 
continuation to a cosmological metric) which  does not depend on 
p-brane charges, i.e. is the same as the metric following by 
reduction from a higher-dimensional `neutral' Schwarzschild 
black hole.

\Date {November 1996}
\noblackbox
\baselineskip 14pt plus 2pt minus 2pt
\lr \dgh {A. Dabholkar, G.W. Gibbons, J. Harvey and F. Ruiz Ruiz,  \np
B340 (1990) 33.}
%
\lr\mon{J.P. Gauntlett, J.A. Harvey and J.T. Liu, \np B409 (1993) 363.}

\lr \CM{ C.G. Callan and  J.M.  Maldacena, 
PUPT-1591,  hep-th/9602043.} 
\lr\SV {A. Strominger and C. Vafa, HUTP-96-A002,  hep-th/9601029.}

\lr\MV {J.C. Breckenridge, R.C. Myers, A.W. Peet  and C. Vafa, HUTP-96-A005,  hep-th/9602065.}

\lr \CT{M. Cveti\v c and  A.A.  Tseytlin, 
\pl { B366} (1996) 95, hep-th/9510097. 
}
\lr \CTT{M. Cveti\v c and  A.A.  Tseytlin, 
IASSNS-HEP-95-102, hep-th/9512031. 
}
\lr\LW{ F. Larsen  and F. Wilczek, 
PUPT-1576,  hep-th/9511064.    }
\lr\TT{A.A. Tseytlin, \mpl A11 (1996) 689,   hep-th/9601177.}
\lr \HT{ G.T. Horowitz and A.A. Tseytlin,  \pr { D51} (1995) 
2896, hep-th/9409021.}
\lr\khu{R. Khuri, \np B387 (1992) 315; \pl B294 (1992) 325.}
\lr\CY{M. Cveti\v c and D. Youm,
 UPR-0672-T, hep-th/9507090; UPR-0675-T, hep-th/9508058; 
  \pl { B359} (1995) 87, 
hep-th/9507160.}

\lr\ght{G.W. Gibbons, G.T. Horowitz and P.K. Townsend, \cqg 12 (1995) 297,
hep-th/9410073.}
\lr\dul{M.J. Duff and J.X. Lu, \np B416 (1994) 301, hep-th/9306052. }
\lr\hst {G.T. Horowitz and A. Strominger, hep-th/9602051.}
\lr\dull{M.J. Duff and J.X. Lu, \pl B273 (1991) 409. }
\lr \guv{R. G\"uven, \pl B276 (1992) 49. }
\lr \gups {S.S. Gupser, I.R.   Klebanov  and A.W. Peet, 
hep-th/9602135.}
\lr \dus { M.J. Duff and  K.S. Stelle, \pl B253 (1991) 113.}

\lr\hos{G.T.~Horowitz and A.~Strominger, Nucl. Phys. { B360}
(1991) 197.}
\lr\teit{R. Nepomechi, \pr D31 (1985) 1921; C. Teitelboim, \pl B167 (1986) 69.}
\lr \duf { M.J. Duff, P.S. Howe, T. Inami and K.S. Stelle, 
\pl B191 (1987) 70. }
\lr\duh {A. Dabholkar and J.A. Harvey, \prl { 63} (1989) 478;
 A. Dabholkar, G.W.   Gibbons, J.A.   Harvey  and F. Ruiz-Ruiz,
\np { B340} (1990) 33. }
\lr\mina{M.J. Duff, J.T. Liu and R. Minasian, 
\np B452 (1995) 261, hep-th/9506126.}
\lr\dvv{R. Dijkgraaf, E. Verlinde and H. Verlinde, hep-th/9603126.}
\lr\gibb{G.W. Gibbons and P.K. Townsend, \prl  71
(1993) 3754, hep-th/9307049.}
\lr\town{P.K. Townsend, hep-th/9512062.}
\lr\kap{D. Kaplan and J. Michelson, hep-th/9510053.}
\lr\hult{
C.M. Hull and P.K. Townsend, Nucl. Phys. { B438} (1995) 109;
P.K. Townsend, Phys. Lett. {B350} (1995) 184;
E. Witten, \np B443 (1995) 85; 
J.H. Schwarz,  \pl B367 (1996) 97, hep-th/9510086, hep-th/9601077;
P.K. Townsend, hep-th/9507048;
M.J. Duff, J.T. Liu and R. Minasian, 
\np B452 (1995) 261, hep-th/9506126; 
K. Becker, M. Becker and A. Strominger, Nucl. Phys. { B456} (1995) 130;
I. Bars and S. Yankielowicz, hep-th/9511098;
P. Ho{\v r}ava and E. Witten, Nucl. Phys. { B460} (1996) 506;
E. Witten, hep-th/9512219.}
\lr\beck{
K. Becker and  M. Becker, hep-th/9602071.}
\lr\aar{
O. Aharony, J. Sonnenschein and S. Yankielowicz, hep-th/9603009.}
\lr\ald{F. Aldabe, hep-th/9603183.}

\lr \ddd{E. Witten, hep-th/9510135;
M. Bershadsky, C. Vafa and V. Sadov, hep-th/9510225;
A. Sen, hep-th/9510229, hep-th/9511026;
C. Vafa, hep-th/9511088;
M. Douglas, hep-th/9512077. }

\lr \gig{G.W. Gibbons, M.J. Green and M.J. Perry, 
hep-th/9511080.}

\lr \dufe{M.J. Duff, S.  Ferrara, R.R. Khuri and 
J. Rahmfeld, \pl B356 (1995) 479,  hep-th/9506057.}

\lr\stp{H. L\" u, C.N. Pope, E. Sezgin and K.S. Stelle, \np B276 (1995)  669, hep-th/9508042.}
\lr \duff { M.J. Duff and J.X. Lu, \np B354 (1991) 141. } 
\lr \pol { J. Polchinski, \prl 75 (1995) 4724,  hep-th/9510017.} 
\lr \iz { J.M. Izquierdo, N.D. Lambert, G. Papadopoulos and 
P.K. Townsend,  \np B460 (1996) 560, hep-th/9508177. }

\lr \US{M. Cveti\v c and  A.A.  Tseytlin, 
\pl {B366} (1996) 95, hep-th/9510097;   hep-th/9512031.  
}
\lr\mast{J. Maldacena and A. Strominger, hep-th/9603060.}
\lr \CY{M. Cveti\v c and D. Youm,
 \pr D53 (1996) 584, hep-th/9507090.  }
 \lr\kall{R. Kallosh, A. Linde, T. Ort\' in, A. Peet and A. van Proeyen, \pr { D}46 (1992) 5278.} 
\lr \grop{R. Sorkin, Phys. Rev. Lett. { 51 } (1983) 87;
D. Gross and M. Perry, Nucl. Phys. { B226} (1983) 29. 
}

\lr \berg{E. Bergshoeff, C. Hull and T. Ortin, 
\np B451 (1995) 547, hep-th/9504081.}

\lr \klts{I.R. Klebanov and A.A. Tseytlin,
 \np B475 (1996) 179,
hep-th/9604166. }
\lr \kltse{I.R. Klebanov and A.A. Tseytlin,
 \np B475 (1996) 179,
hep-th/9607107. }
\lr \cvets{ M. Cveti\v c  and A.A. Tseytlin, \np B478 (1996) 181, 
hep-th/9606033. }

\lr \tset  {A.A.  Tseytlin,  \np B475 (1996)
 149, hep-th/9604035.}
 
\lr \horts{G.T. Horowitz and A.A.  Tseytlin, \pr D51 (1994) 3351,
hep-th/9408040.}

\lr \horow{  J.H. Horne, G.T. Horowitz and 
 A.R. Steif, \prl 68 (1992) 568;
 G. Horowitz, in {\it String Theory and Quantum Gravity '92}, 
 ed. by J. Harvey et al.  (World Scientific, Singapore, 1993). }

\lr\beh {K. Behrndt, E. Bergshoeff and B. Janssen, hep-th/9604168.}

\lr \ruts{J.G. Russo  and A.A. Tseytlin,
hep-th/9611047. }

\lr\lar{F. Larsen and F. Wilczek, hep-th/9610252.}

\lr\behrn{K. Behrndt and S. F\"orste, \np B430 (1994) 441,
hep-th/9403179.}
 \lr\popp{R. Poppe and S. Schwager, hep-th/9610166.}
 
 \lr\mark{V.P. Frolov,  M.A. Markov and  V.F. Mukhanov, 
\pl  B216 (1989) 272; \pr D41 (1990) 383.}

 \lr\dlp{ M.J. Duff, H. L\"u and  C.N. Pope,
  \pl  B382 (1996) 73, hep-th/960405. }

\lr\hos{G.T.~Horowitz and A.~Strominger, \np { B360}
(1991) 197.}
\lr \guven{R. G\"uven, \pl B276 (1992) 49. }
\lr \gibbon{G.W. Gibbons, \np B207 (1982) 337. }

\lr \KKM{ D.  Gross and M.  Perry, \np B226 (1983) 29;
R.   Sorkin, \prl 51 (1983) 87.} 

\lr\hms{ G. Horowitz, J. Maldacena and A. Strominger, \pl B383 (1996)
151,  hep-th/9603109.}
\lr \cyo{ M. Cveti\v c  and D. Youm,  \np B476 (1996) 118, 
hep-th/9603100.}
\lr \duflu { M.J. Duff and J.X. Lu, \np B354 (1991) 141. } 

\lr \chs{C.G. Callan, J.A. Harvey and A. Strominger, 
\np { B359 } (1991)  611.}
\lr \duklu { M.J. Duff, R.R. Khuri  and J.X. Lu, 
Phys. Rept. 259 (1995) 213,  
hep-th/9412184.} 
\lr \gaunt{J.P.~Gauntlett, D.A.~Kastor and J.~Traschen, 
hep-th/9604179.}

\lr\rust{J.G. Russo and A.A. Tseytlin, hep-th/9611047.}
\lr \john {J.H.  Schwarz, \pl B360 (1995) 13,
hep-th/9508143.}
\lr \papd{G. Papadopoulos and P.K. Townsend, \pl B380
 (1996) 273, hep-th/9603087.}
\lr\jch {J. Polchinski, S. Chaudhuri and C.V. Johnson, 
hep-th/9602052.}

\lr\wald{D. Waldram, \pr D47 (1993) 2528.}
\lr \HT{ G.T. Horowitz and A.A. Tseytlin,  \pr { D51} (1995) 
2896, hep-th/9409021.}
\lr\TT{A.A. Tseytlin, \mpl A11 (1996) 689,   hep-th/9601177.}

Suppose one starts   with a  configuration of 
intersecting black 
p-branes  in $D=11$ supergravity,  or the corresponding
configuration of R-R and NS-NS p-branes 
 in $D=10$ theory,     such that it  becomes  BPS saturated 
in the extremal limit \refs{\papd,\tset,\klts,\gaunt}.
 Let the total  number of  internal directions 
 of intersecting branes (or the dimension of the corresponding
 anisotropic    brane \guven) 
 be $p$  so that  the number 
 of remaining   `transverse' spatial and time 
  directions is $D$ (with $D+p = 11$ or $10$). 
 Wrapping the branes around an internal p-torus  and 
dimensionally reducing 
one finds the following 
spherically symmetric static  black hole (Einstein-frame)  
metric in $D$ dimensions  \refs{\cvets,\kltse}\foot{Some 
special  cases were discussed, 
e.g., in \refs{\hos,\guven,\duklu,\dlp,\cyo,\hms} (see \cvets\
for a more complete list of references).}
\eqn\met{
ds^2_D=   h^{1\ov D-2} (r)  \big[ - h\inv (r) f(r)  dt^2+   
f\inv (r) dr^2+r^2d\Omega^2_{D-2} \big]
\  ,}
$$
h(r) = \prod^N_{i=1}  H_i(r) \ , \ \ \  \ \ 
\  f(r) = 1 -  { 2\mu \ov r^{D-3} } \ ,\ \ 
\ \ \ \ 
H_i(r) = 1 + { \Q_i \ov r^{D-3} }  \ ,  
   $$
where $ \Q_i = \sqrt {Q_i^2 + \m^2} -\m$  depend on 
the brane  charge parameters  $Q_i$ and  single  
 non-extremality
 parameter $\mu$. 
\  $f$ is the standard Schwarzschild function 
and $h$ is  the product of `harmonic' functions,\foot{Non-extremal
solutions do not have, of course, 
 static multicenter generalisations.} 
 one for each 
brane  in the intersection  
(counting also the linear momentum  of a possible boost
 along  a common string 
 as a separate   `charge').

Since this black hole  is static, we can formally
reduce further in  time  direction,    down to $D-1$ dimensions.
This reduction may be motivated   by  considering 
a `cosmological' solution (as, e.g., in  
\refs{\behrn,\popp,\lar}) 
related to the background \met\ by the analytic continuation 
$t = i y_0$,\ $r= i\tau$ with compact $y_0$.\foot{In the case 
when $D$ is even  so that the power  of $r$  in
$H_i$  is  odd  one  needs to  rotate also $\Q_i$ 
to get a real metric. Other background fields (i.e. the 3-tensor  
if  the continuation is done directly at the level of p-brane 
configuration  in $D=11$) 
 also remain real.}
Then $ \int d^{D} x \sqrt {g_D} R_D + ... \to 
\int d^{D-1} x   V(x)  
 \sqrt{ g_{D-1}} R_{D-1} + ..., 
 \ \ V= (h^{3-D\ov D-2} f)^{1/2} $,  so that 
 transforming to the Einstein frame,  
$g_{D-1} = V^{2\ov D-3 } g^{(E)}_{D-1}$, 
one finishes with\foot{Notice the similarity  between 
 $f$-dependence of 
 this metric 
  and  $h$-dependence  of \met\ 
dimensionally reduced to $D-1$ dimensions 
along a spatial direction (by taking a periodic array of black holes,
etc.). The latter metric turns out to be  
 again   \met\   with $D$ replaced by 
$D-1$.}
\eqn\mett{
ds^2_{D-1} =  
 f^{1\ov D-3} (r)  \big[   
f\inv (r) dr^2+r^2d\Omega^2_{D-2} \big]
\  . }
This is exactly  the same metric 
 that  follows  simply  from the 
\Sc\ metric  in the space  with $p$ isometries, 
i.e. from  the trivial `{\it neutral}  black $p$-brane',   
\eqn\mee{
ds^2_{D+p} =  
  -f(r)  dt^2 + dy_1^2 + ... + dy_p^2 +  
f\inv (r) dr^2+r^2d\Omega^2_{D-2}  \ , }
after  one reduces it along all  the  $p+1$  isometric 
(p-brane world-volume)  directions. 
The  structure of  \met\ is such that upon reduction in the time
direction the dependence on the function $h$, i.e. on the charges
of the  branes, completely  disappears!

In the special case of $D=5$,   which is of 
interest in connection with  cosmology in  four dimensions, 
 an equivalent  observation was recently  made in \lar.\foot{See also 
  \popp\  where  the case of the reduction of a single 5-brane 
  was discussed.} 
 This implies
 that \mee\ or related cosmological metric  can be  embedded
 into  any of  higher-dimensional black p-brane  configurations 
 which  correspond to \met. 
 
Our aim below  will be  to give a simple  explanation
 for  this
universal 
decoupling of the  brane charges 
  upon reduction in all world-volume directions. In the process, 
   we
  shall  review and clarify 
 the general structure of the composite black brane 
configurations.

The main message  is that the \Sc\  or the `neutral'
black p-brane  solution \mee\ is,  essentially,   the `back-bone'
of all such  intersecting   black p-brane backgrounds in $D=10$ or $D=11$. 
All $D=11$ solutions in \refs{\cvets} and some of their 
 generalisations \rust\  
 can be constructed from \mee\ 
 with $p \geq 2$
by applying  combinations of the following 
transformations:
(a) a  Lorentz  boost   of \mee\ along one of the
isometric directions $y_i$; (b) a
transformation  `trading' off-diagonal components of the metric 
for the antisymmetric 3-tensor  background, 
  which is just the standard 
$T$-duality  from the point of view of the  reduced 
10-dimensional theory\foot{In the case of at least two spatial 
isometries 
the $D=11$ action 
 has the same  $T$-duality  symmetry as  the  $D=10$ action  in
  the case of at least one spatial isometry 
  (we are ignoring the Kaluza-Klein modes, i.e. 
  assume that one deals  essentially  with  
  $D \leq 8$ dimensionally reduced actions \berg).
  We assume that the time direction is always an isometry, i.e.
  the total number of isometries in $D=11$ is $\geq 3$.}; 
  (c) a linear  coordinate transformation  `mixing'
 two  of the spatial isometric coordinates, which 
 corresponds  to an $SL(2)$ transformation  of  type IIB 
 10-dimensional  
 theory \refs{\john,\berg}.

 We shall  use    
  the familiar $D=10$  supergravity (string-theory) 
    language 
     since   the discussion of   $D=11$ solutions with $p \geq 2$
 isometries  can be also rephrased in  it. 
 The first two  basic generating transformations 
 correspond to   $O(p+1,p+1)$ duality.
 The  central starting point is  the \Sc\ metric 
  \mee\   
 boosted \gibbon\ to  some finite momentum  along, e.g., $y_1$ direction,  
 \eqn\wave{
ds^2 =  -f(r) dt'^2  + dy'^2_1  + dy_2^2  + ...+ dy^2_p + 
f\inv (r) dr^2 +  r^2 d\Omega_{D-2}^2  }
$$
=  - dt^2 + dy^2_1   + ...+ dy^2_p + 
 (f(r) -1)  (\cosh \beta\ dt - \sinh \beta\ dy_1)^2
 + f\inv (r) dr^2 +  r^2 d\Omega_{D-2}^2  \ ,  $$
 where 
 \eqn\boos{
 t'= \cosh \beta \ t - \sinh \beta\ y_1\ ,  \ \ \ \ \ 
\ y'_1= - \sinh \b\ t + \cosh \beta\ y _1 \ .  }
 The background \wave\  becomes a gravitational wave  
 in the extremal limit $\m\to 0,$ $
 \b\to \infty,$  $Q=\m \sinh 2\b =$fixed.
 $T$-duality  applied in $y_1$ direction then leads \horow\ to a  
 new solution (now already with 
  non-trivial dilaton and $B_{\m\n}$) 
 --  the non-extremal generalisation \hos\ of the 
 fundamental string \dgh, or charged black string.\foot{More precisely, for $p > 1$ 
 we get a string
 `smeared' in other $p-1$ isometric directions,  
or,  in extremal limit, as a  periodic array of strings.
Equivalently, this background can be interpreted as a black p-brane 
with a  charge along $y_1$ string  direction only.}
The  $SL(2)$  duality of type IIB theory 
then  produces the  black  R-R string
 solution. Further $T$-dualities \berg\ 
 generate various  black R-R p-brane 
 backgrounds \hos, in particular, the  5-brane of type 
 IIB theory. 
 Applying the 
 $SL(2)$ duality   then leads to the solitonic NS-NS 5-brane, 
 thus effectively
 performing  the `electro-magnetic' duality 
 transformation between 
 the  fundamental string and the 
  solitonic 5-brane \duklu. 
 The $D=10$ black $2$-brane and $5$-brane  backgrounds 
 can then be lifted to $D=11$,  
 thus explaining the structure of the resulting  non-extremal 
 $D=11$ solutions \refs{\guven}   in terms 
 of \wave\ and lower-dimensional dualities. 
 
 To  obtain  the explicit form of all 
 single black brane  backgrounds one,  therefore, 
  never
  needs to solve 
 the  supergravity equations of motion: the only input is the
 Schwarzschild solution \mee\ (in its coordinate-transformed 
 form 
 \wave)  and $T$ and $SL(2)$ symmetries of the
 supergravity actions (or $U$-duality  symmetry 
 of the $D=8$  action \berg). 
 To construct all extremal solutions it is sufficient to start
 with the 0-brane background (with some number $p$ 
 of additional isometries)
 and apply $O(p+1,p+1)$ and $SL(2)$ dualities.
 
 Having found a single black brane background, one may then repeat 
 the  procedure,   by using it,  instead of \wave, 
  as a starting point.
 Appropriately boosting and applying  the 
  duality transformations 
(and `smearing'  when necessary in internal directions)
one can then  generate 
backgrounds describing  intersections 
of pairs of  branes.\foot{For pairs of intersecting R-R branes 
this  can be done simply  by making formal $T$-duality 
transformations 
 as was demonstrated in the extremal case in \beh.} 
  Each `boost+duality' cycle
  introduces  a new  charge or  function $H_i$ into the
 intersecting brane solution (and thus into 
 the corresponding 
  black hole metric
 \met). 
 This essentially explains  the 
 applicability of the `harmonic function rule' \tset\
 (or its non-extremal version  \cvets) 
 to  the construction of such backgrounds.

 Let us illustrate this  construction more explicitly on some 
 basic examples
 of backgrounds corresponding 
 to threshold and non-threshold bound states of
 branes. 
  Having found  the charged black string  by 
 T-duality from  boosted Schwarzschild (neutral string) background, 
 we can finitely boost it  
 in the longitudinal direction 
 (the  string metric is  not boost-invariant in the non-extremal case).
 The resulting
  solution  is parametrised by two charges (two
 harmonic functions in the extremal limit) 
 and represents  a wound string with a 
 linear momentum flow along it \refs{\wald,\HT},  or it  can be  viewed  as a
 combination of a fundamental string and a wave, $1_{NS} + \up$.  
 All other possible
 (threshold or non-threshold) configurations of  intersecting  branes
 with {\it two}  charges  can  now be  obtained  from 
 $1_{NS} + \up$ 
 by various sequences of dualities.
 For example, $SL(2)$ leads to R-R string with momentum, $1_{R} + \up$, 
 and then $T$-duality in the orthogonal isometric  directions  produces
 other longitudinally boosted p-brane backgrounds, e.g., 
   $4 + \up$, $5_{R}  + \up$ and $6+ \up$. 
    $T$-duality in the direction of 
 the momentum flow then gives $3\perp 1_{NS}$, $4 \perp 1_{NS}$,
   and    $5_{R} \perp 1_{NS}$.  $SL(2)$ transforms $5_{R}  + \up$
   into $5_{NS}  + \up$,    leading,  after $T$-duality, 
    to the explicit form 
   of  non-extremal  solitonic 5-brane plus fundamental 
   string  $5_{NS}  + 1_{NS}$ background (and,   further,  to 
   $5_{R}  + 1_{R}$ after $SL(2)$ duality, \  $4+0$ after $T$-duality,  etc.). 
 Alternatively, applying $SL(2)$ rotation to the  black string 
 background one finds the non-extremal version of the string-string bound
 state, $1_{NS} + 1_R$ \john. Other non-threshold bound state configurations 
 with two charges are related to $1_{NS} + 1_R$  by the dualities:  
 $T$-duality gives $3 + 1_{NS}$, $SL(2)$ transforms this into 
 $3 + 1_{R}$,  $T$-duality  along the string direction  
  produces  $2+0$  configuration, etc.
 The explicit form of these  backgrounds was  given  in \rust.
 
 To add  a  third charge (a third brane in  the  intersection)
 one is to boost the two-brane configuration. For example,    boosting  
 $5_{NS} + 1_{NS}$ along the string direction we  obtain 
 $5_{NS} + 1_{NS}+ \up$ and thus  also the  dual backgrounds 
 $5_{R} + 1_{R}+ \up$, $4 + 0 \perp 1$, etc.  Compactification then leads 
  to the three-charge $D=5$ black hole.\foot{This construction of 
  the   non-extremal version of the black hole corresponding to 
  $5_{NS} + 1_{NS}+ \up$ solution of  \TT\
  is different from  the one in \hms\  
  in that here the starting point is not the equal-charge Reissner-Nordstr\"om
 type solution but just the  Schwarzschild solution 
  (or neutral 5-brane) and all  three  charges are added by boosts interchanged with 
  duality transformations.}

Notice  that one never needs to decide 
which  combinations of branes are possible  -- they are 
 automatically selected  by  the duality transformations.\foot{This is true
  not only for combinations of R-R branes where there is a 
 microscopic  D-brane explanation for $T$-duality relations \jch\ but also 
 for `mixed' NS-NS and R-R combinations.}
 As the duality transformations preserve supersymmetry of the extremal
 solutions, one may say that the origin 
 of  the  1/2 supersymmetry of the  single  branes 
 (or non-threshold bound states related to some of 
 them by  $SL(2)$  rotations) 
 is in supersymmetry of a  wave or  $T$-dual fundamental string background, 
 while the reason for the  1/4 supersymmetry of the
  threshold combinations of two branes
 is in the 1/4  supersymmetry of the 
  fundamental string with momentum, etc.
 Another remark is that the  procedure described above
 implies that the resulting non-extremal
solutions have just one 
non-extremality parameter -- the mass of the original Schwarschild solution, 
i.e. they are {\it one}-parameter `deformations' of the supersymmetric BPS backgrounds.

 Now, why  the above construction 
  guarantees that  if  one reduces in all internal
 brane directions  {\it and}  time  direction  one 
  should  get the metric \mett\
 with no trace of the $H_i$-functions? There are two simple reasons.
 First reason  is that if one reduces also in time direction 
 (or in 
 its compact analytic continuation)  
 the boost \boos\  becomes 
 simply an `internal' coordinate transformation which does not 
 influence  
  the dimensionally reduced metric. Second    is the 
 duality invariance of the  reduced  Einstein-frame metric.

This   explanation of  
  why the intersecting brane metric or
  \met\ always reduces to \mett\   may have  some implications
    for  an  
 interesting recent 
discussion \lar\  of possible string-theory 
resolution of  cosmological singularities.
As far as the Einstein-frame   metric is concerned,
  all 
`exotic'    higher-dimensional
embeddings of \mett\  which are  related to 
(R-R, NS-NS, or `mixed')  p-brane configurations
   should  be   considered on an equal footing
with  the  simplest `neutral' Schwarzschild 
embedding \mee. It remains to be seen  
whether the presence of additional
matter fields  makes some of more complicated embeddings 
more  promising.\foot{In particular, there is
 a question of whether 
 $D$-branes  should  have any special 
  importance in this context   
(given also that  the non-extremality of the solutions 
is  essential  here:  for $\mu=0$ \mett\ becomes simply a
flat space).}

\smallskip

 This work was partially supported by PPARC,
 the European
Commission TMR programme ERBFMRX-CT96-0045 and 
 NATO grant CRG 940870.

\vfill\eject
\listrefs
\end